# High-Pressure Torsion-Induced Transformation of Adenosine Monophosphate: Insights into Prebiotic Chemistry of RNA by Astronomical Impacts

Kaveh Edalati[1,2,3,]*, Jacqueline Hidalgo-Jiménez[1,2], Thanh Tam Nguyen[1,3]

[1] WPI, International Institute for Carbon Neutral Energy Research (WPI-I2CNER), Kyushu University, Fukuoka 819-0395, Japan
[2] Mitsui Chemicals, Inc. - Carbon Neutral Research Center (MCI-CNRC), Kyushu University, Fukuoka 819-0395, Japan
[3] Graduate School of Integrated Frontier Sciences, Department of Automotive Science, Kyushu University, Fukuoka 819-0395, Japan

The origin of life is yet a compelling scientific mystery that has sometimes been attributed to high-pressure impacts by small solar system bodies such as comets, meteoroids, asteroids, and transitional objects. High-pressure torsion (HPT) is an innovative method with which to simulate the extreme conditions of astronomical impacts and offers insights relevant to prebiotic chemistry. In the present study, we investigated the polymerization and stability of adenosine monophosphate (AMP), a key precursor to ribonucleic acid (RNA), in dry and hydrated conditions (10 wt% water) under 6 GPa at ambient and boiling water temperatures. Comprehensive analyses with the use of X-ray diffraction, Raman spectroscopy, Fourier-transform infrared spectroscopy, nuclear magnetic resonance, scanning electron microscopy, and matrix-assisted laser desorption/ionization time-of-flight mass spectrometry revealed no evidence of polymerization, while AMP partly transformed to other organic compounds such as nucleobase-derived fragments of adenine, phosphoribose fragments, dehydrated adenosine, protonated adenosine, and oxidized adenosine. The torque measurements during HPT further highlight the mechanical behavior of AMP under extreme conditions. These findings suggest that, while HPT under the conditions tested does not facilitate polymerization, the formation of various compounds from AMP confirms the significance of astronomical impacts on the prebiotic chemistry of RNA on early Earth.

*Corresponding author: Kaveh Edalati (E-mail: kaveh.edalati@kyudai.jp; Tel: +80-92-802-6744)

# 1. Introduction

The origin of ribonucleic acid (RNA) from simpler organic molecules, such as adenosine monophosphate (AMP), guanosine monophosphate (GMP), thymidine monophosphate (TMP), and cytidine monophosphate (CMP), is an essential step in the emergence of life (Patel et al., 2015). Astronomical impacts caused by comets, meteoroids, asteroids, and transitional objects colliding with early Earth are hypothesized to have created extreme conditions favorable for chemical evolution (Rotelli et al., 2016). These conditions, characterized by high pressures, temperatures, and strains, are thought to drive molecular transformations that may lead to the synthesis of biologically relevant polymers or other organic compounds essential for life (Blank et al., 2001). AMP with a composition of $C_{10}H_{14}N_5O_7P$, a nucleotide monomer of RNA with the molecular structure shown in Fig. 1a, represents one of the key building blocks for life and offers unique insights into prebiotic chemistry (Kitadai and Maruyama, 2018).

The detection of extraterrestrial organic molecules such as amino acidss, nucleotides, and sugars in meteorites and comets, suggests that extraterrestrial sources may have contributed to the prebiotic inventory on early Earth (Martins et al., 2008; Glavin et al., 2010). For example, the Murchison meteorite contains various amino acids (Martins et al., 2008), while recent studies have identified nucleobases in other carbonaceous chondrites (Glavin et al., 2010). Moreover, theoretical simulations and shock experiments indicate that impact-generated conditions can foster chemical evolution that enables the synthesis of amino acids and nitrogen-containing polycyclic aromatic hydrocarbons (Martins et al., 2013; Kroonblawd et al., 2019). High-pressure and high-temperature conditions mimic those of impact events and have been proposed as key environments for prebiotic chemistry, while recent studies have suggested that the role of strain during such impacts cannot be neglected (Edalati et al., 2022b; Edalati et al., 2024b). It should be noted that strain is defined as the deformation of materials and is quantified as a unitless parameter by comparing the dimensions after deformation with the dimensions before deformation (Edalati et al., 2022a).

High-pressure torsion (HPT), as illustrated in Fig. 1b, is an experimental technique that combines high pressure with strain (Bridgman, 1935; Zhilyaev and Langdon, 2008; Edalati and Horita, 2016) and has recently been used for simulating the combined physical stresses during astronomical impacts (Edalati et al., 2022b; Edalati et al., 2024b). Unlike conventional high-pressure methods, such as shock experiments, HPT applies strain in addition to compression, which enables a better approximation of impact dynamics. Previous studies have examined the stability and transformation of amino acids, such as glycine (Edalati et al., 2022b), serine (Edalati et al., 2024b), and glutamic acid (Edalati et al., 2024b by using the HPT method, but polymerization was not observed, even though alcohol has been detected by the transformation of glycine. In spite of these studies, there have been no known attempts to investigate the polymerization of the building precursors of RNA when using the HPT method.

The astrobiological significance of nucleotide stability and polymerization is profound with regard to understanding the origin of life (Patel et al., 2015; Kitadai and Maruyama, 2018). It is hypothesized that nucleotide precursors could survive, potentially polymerize during impact events, and, thus, contribute to the formation of RNA (Benner et al., 2012; Furukawa et al., 2019). However, experimental evidence of polymerization is lacking. Understanding the behavior of AMP under such conditions could illuminate pathways for the prebiotic synthesis of nucleic acids and provide a critical link in the puzzle of life's origins (Patel et al., 2015; Kitadai and Maruyama, 2018). The present study, thus, evaluates the stability and reactivity of AMP under HPT conditions and simulates dry and hydrated environments at ambient and boiling water temperatures (300 K and 373 K) to reflect varying early Earth conditions. These findings offer insight into prebiotic chemistry and highlight the capabilities of HPT as a tool for the study of molecular transformations under extreme astronomical impact conditions.

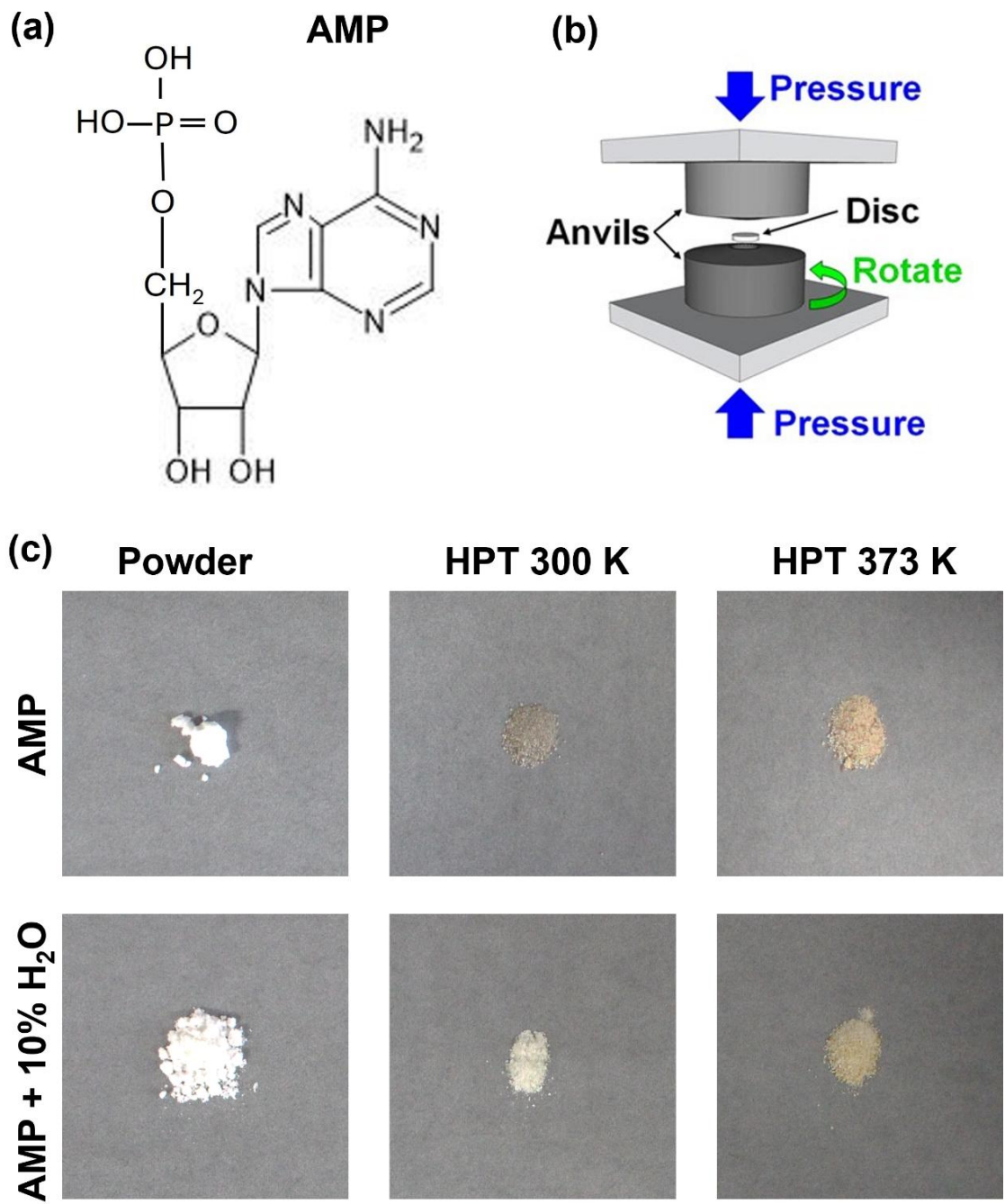

Fig. 1. Schematic representation of (a) adenosine monophosphate and (b) high-pressure torsion (Edalati et al., 2022b). (c) Appearance of adenosine monophosphate in dry and hydrated conditions before and after treatment through high-pressure torsion at ambient and boiling water temperatures.

## 2. Methods
### 2.1. Sample Preparation

Powder of adenosine 5'-monophosphate with >97% purity was purchased from Sigma-Aldrich, USA. Samples were prepared in two forms: dry powder and hydrated powder, the latter obtained by mixing the dry powder with 10 wt% deionized water for 10 minutes. Hydrated samples were continuously homogenized by manual mixing before compression to ensure uniform distribution of water. The dry and hydrated samples were used to simulate conditions with varying moisture levels, which reflects the diversity of environments on early Earth. This approach allows for the evaluation of the role of hydration in facilitating molecular transformations during astronomical impacts. Both dry and hydrated AMP samples were compressed into disc shapes (5 mm radius, 0.8 mm thickness) under 300 MPa by using a cylindrical die and a plunger. The prepared discs were subsequently subjected to HPT processing (Zhilyaev and Langdon, 2008; Edalati and Horita, 2016).

### 2.2. High-Pressure Torsion

HPT experiments were performed by using an apparatus equipped with two cylindrical anvils. The setup applied a static pressure of 6 GPa to AMP, while shear deformation was induced by turning the lower HPT anvil at a rate of 1 rotation per minute for three rotations. This resulted in a maximum shear strain of approximately 120 (dimensionless) at the periphery of the disc sample. Processing was conducted at ambient (300 K) and boiling water (373 K) temperatures to explore the temperature-dependent effects of HPT on AMP. For the samples processed at 373 K, the rotation started at 371 K and ended at 375 K. The mixture of high pressure and shear deformation in HPT replicates the extreme conditions associated with astronomical impacts, including high mechanical pressure, strain, and thermal effects, although the quantity of these parameters may not represent the exact conditions during astronomical impacts (Martins et al., 2013; Kroonblawd et al., 2019).

### 2.3. Characterization

Materials were analyzed by using various methods, as described in this section.

(1) Based on Bragg's Law, structural analysis was conducted by X-ray diffraction (XRD) and the use of a Rigaku Smart Lab X-ray diffractometer with Cu Kα radiation, with a step of 0.05° and a speed of 2 °/min. XRD patterns were collected for both pre- and post-HPT-treated samples to detect phase transformations or crystallinity changes.

(2) A 785 nm laser source was used to obtain Raman spectra with a Renishaw inVia Raman Microscope with 50x magnification, with a focus on vibrational modes associated with the molecular structure of AMP. The selected laser was used to minimize any fluorescence that might interfere with the analysis. For this analysis, the samples did not require any special preparation; the resulting powders were flattened between two square glass slides.

(3) Fourier-transform infrared spectrometer (FTIR) Thermo Scientific™ Nicolet iN10 MX was employed for the identification of changes in functional groups, particularly those associated with polymerization.

(4) Proton $^1$H, carbon $^{13}$C, and phosphorous $^{31}$P nuclear magnetic resonance (NMR) spectroscopy was carried out by using a 600 MHz Bruker AvanceIII600 spectrometer. Nearly 20 mg of materials were added to 0.5 ml of dimethyl sulfoxide (DMSO). The samples were dissolved in DMSO because (i) DMSO effectively dissolves AMP and its transformation products, which ensures uniform distribution for accurate spectral measurement, and (ii) DMSO has low interference with NMR signals, which enables reliable identification of molecular peaks. After the dissolution of material in DMSO, the liquid phase was poured into an NMR tube, and the resonance was monitored at 600, 151, and 243 MHz for $^1$H, $^{13}$C, and $^{31}$P spectra, respectively.

(5) Morphological changes were visualized by using scanning electron microscopy (SEM) with 15 kV acceleration voltage, a light-emitting diode (LED) detector, and a magnification of 500X in a JEOL JSM-7900F scanning electron microscope. SEM samples were prepared by dispersing powders on carbon tape, followed by coating with gold using a magnetron sputtering system. Images were captured for both dry and hydrated samples before and after HPT treatment.

(6) Molecular mass distributions were analyzed by a matrix-assisted laser desorption/ionization time-of-flight (MALDI-TOF) mass spectrometer, Bruker Microflex LT, for detecting any polymerization products or molecular fragments. To prepare the samples for MALDI-TOF mass spectroscopy, 5 mg of each sample were diluted in 0.5 ml of a matrix solution that consisted of acetonitrile with 0.1 vol% trifluoro acetic acid (8.5 ml from Fujifilm Wako Pure Chemical, Japan), deionized water (1.5 ml), and α-cyano-4-hydroxycinnamic acid (5 mg from Sigma-Aldrich, U.S.A.) in separated containers following a procedure described elsewhere (Gogichaeva and Alterman, 2019). Subsequently, the solution was placed in the ultrasonic bath and cooled down with iced water for 5 min. Then, a 0.02 ml drop of these mixtures was deposited over the sample holder and examined by MALDI-TOF mass spectroscopy. The analysis was performed soon after preparing the samples to avoid any changes over time.

(7) Torque was measured in situ during HPT processing to evaluate shear stress ($\tau$) versus shear strain ($\gamma$) evolution. Results were compared to reference materials, aluminum and copper, to understand the mechanical response of AMP. The shear stress and strain were estimated by using the following equations (Edalati et al., 2024b).

$$\tau = \frac{3q}{2\pi R^3} \quad (1)$$

$$\gamma = \frac{2\pi R N}{h} \quad (2)$$

where $q$ is the in situ measured torque, $R$ is the radius of the disc, $N$ is the lower anvil rotation numbers, and $h$ is the height of the sample.

## 3. Results

Photographs of the materials are shown in Fig. 1c. While the initial AMP powder in both dry and hydrated forms has a white color, its color changes to dark after HPT processing. Such color changes usually indicate structural changes and/or chemical transformations (Sasmal and Pal, 2021).

XRD analyses (Fig. 2) confirmed the partial retention of the crystalline structure of AMP after HPT at both 300 K and 373 K. However, the appearance of a broad peak from 15º to 30º indicates that amorphization occurred by HPT processing, and such amorphization is more significant after HPT processing at an elevated temperature of 373 K. A closer examination of the XRD profiles at a magnified view (Fig. 2c) show peak shifts to lower diffraction angles, which indicates a lattice expansion occurred, despite high applied pressure during HPT processing. Such lattice expansions are generally due to the formation of point defects such as vacancies, as reported for various HPT-processed materials (Oberdorfer et al., 2010; Čížek et al., 2019).

Raman spectra (Fig. 3) revealed that the peaks of the starting powders in both dry and hydrated forms fit well with the reference profiles in the literature for AMP (Rimai et al., 1969; Kundu et al., 2009). No additional peaks indicative of polymerization were detected after HPT processing. However, HPT-processed samples exhibited an increase in background intensity and the disappearance of some peaks, which indicates structural destruction and lattice defect formation after HPT processing (Kitajima, 1997).

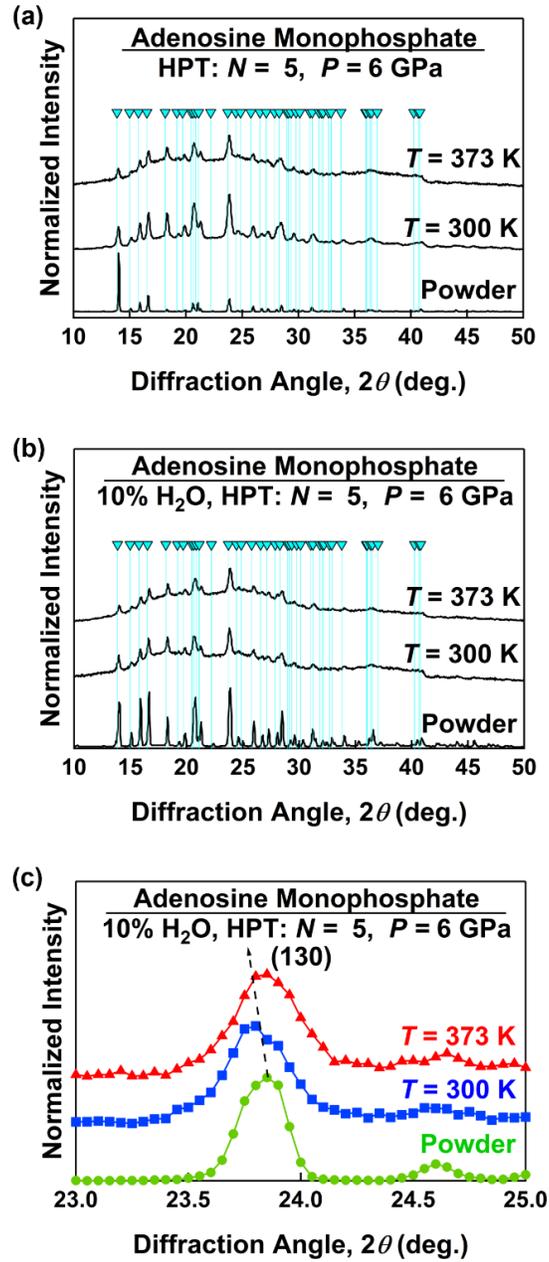

Fig. 2. X-ray diffraction patterns of adenosine monophosphate in (a) dry and (b) hydrated conditions before and after treatment through high-pressure torsion at ambient and boiling water temperatures, where (c) shows magnified view of (130) atomic plane for hydrated samples.

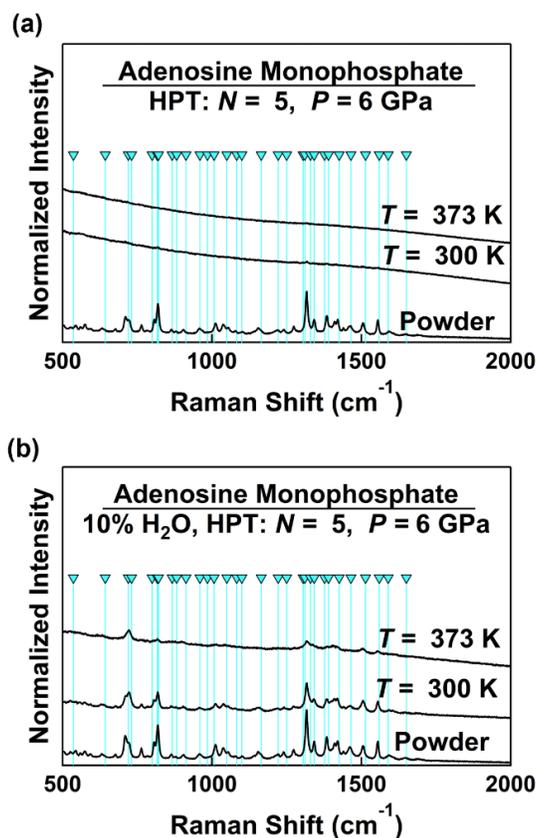

Fig. 3. Raman spectroscopy patterns of adenosine monophosphate in (a) dry and (b) hydrated conditions before and after treatment through high-pressure torsion at ambient and boiling water temperatures.

FTIR spectra of AMP in (a) dry and (b) hydrated conditions Fig. 4 revealed no significant changes in functional group vibrations. The absence of new absorption bands suggests a lack of significant polymerization. The FTIR spectra shown in Fig. 4 are in good agreement with the reported spectra in the literature for AMP (Wang et al, 2022; Rakshit et al., 2022).

NMR spectra for (a,b) $^1$H, (c,d) $^{13}$C, and (e,f) $^{31}$P are illustrated for (a, c, e) dry condition and (b, d, f) hydrated condition in Fig. 5. NMR spectra showed characteristic peaks for AMP (Arranz-Mascaros et al., 2011; Cho and Kim, 2022), with no clear evidence for polymerization. The slight differences between the position of AMP peaks compared to the literature can be due to pH, solvent, and temperature effects. After HPT processing, changes in the shape of spectra, relative peak intensities, and slight peak shifts indicate the occurrence of chemical transformations. Detailed analysis of NMR signals and a comparison of them with the reported data in the literature confirmed the presence of adenosine (Chemical Book, 2025) and adenine (Qi et al., 2021), which is an indication of the decomposition of AMP into smaller molecules rather than its polymerization. Moreover, $^{31}$P NMR spectra provided no evidence for adenosine diphosphate (ADP) or triphosphate (ATP) formation (Arranz-Mascaros et al., 2011; Cade-Menun, 2015; Nagana Gowda et al., 2016).

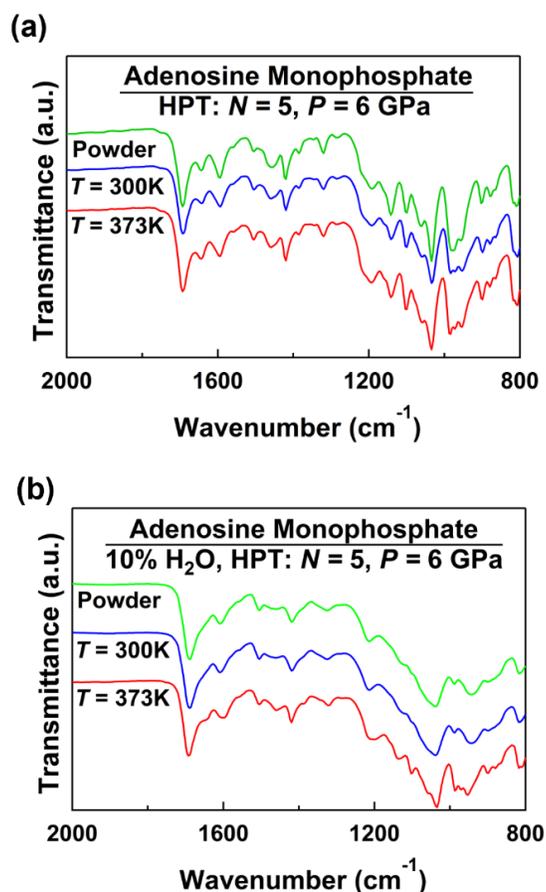

Fig. 4. Fourier-transform infrared spectra of adenosine monophosphate in (a) dry and (b) hydrated conditions before and after treatment through high-pressure torsion at ambient and boiling water temperatures.

MALDI-TOF mass spectroscopy analysis for the initial unprocessed powder (Fig. 6) revealed the presence of AMP at a mass-to-charge ratio of $m/z = 347.5$, which agrees well with previous publications (Liu et al., 2006). The intensity of the AMP peak decreased after HPT processing at 300 K, and no AMP was detected after processing at 373 K. New peaks at different $m/z$ values appeared: $m/z = 135.1$ corresponds to the nucleobase-derived fragment adenine $C_5H_5N_5$, $m/z = 189.3$ and 211.3 correspond to the phosphoribose fragment $C_5H_9O_5P$, $m/z = 249.5$ corresponds to dehydrated adenosine $C_{10}H_{11}N_5O_3$, $m/z = 267.5$ corresponds to protonated adenosine $C_{10}H_{14}N_5O_4^+$, and $m/z = 283.2$ corresponds to an oxidized form of adenosine $C_{10}H_{13}N_5O_5$ (Fu et al., 2000; Liu et al., 2006; Strzelecka et al., 2017).

Among these products, the formation of the nucleobase-derived fragment adenine and protonated adenosine was also confirmed with NMR. The signal for oxidized adenosine superimposes with the signal for guanosine (Yamagaki and Nobuhara, 2025). Although it may be expected that extreme conditions of high pressure and strain might overcome the energy barriers for guanosine formation, it is hypothesized that oxidized adenosine formed at 373 K due to the contact of the materials with atmospheric air during the high-temperature process. Nevertheless, no higher-order polymers were observed (Fig. 6) after HPT processing, which confirms the absence of polymerization and the decomposition in AMP under the tested conditions, similar to the observations for glycine (Edalati et al., 2022b).

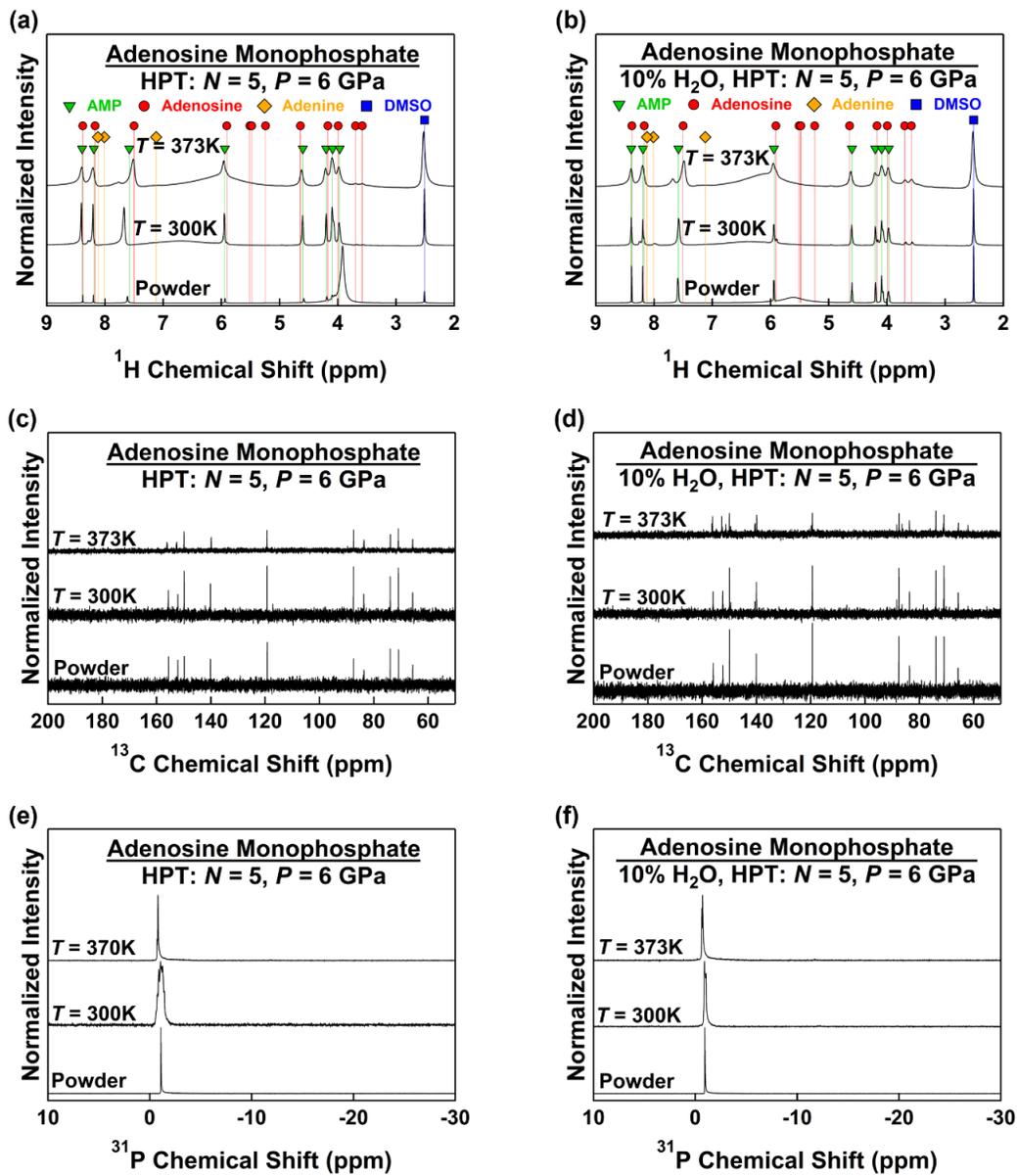

Fig. 5. Nuclear magnetic resonance spectra for (a, b) $^1$H, (c, d) $^{13}$C and (e, f) $^{31}$P of adenosine monophosphate in (a, c, e) dry and (b, d, f) hydrated conditions prior and after treatment through high-pressure torsion at ambient and boiling water temperatures.

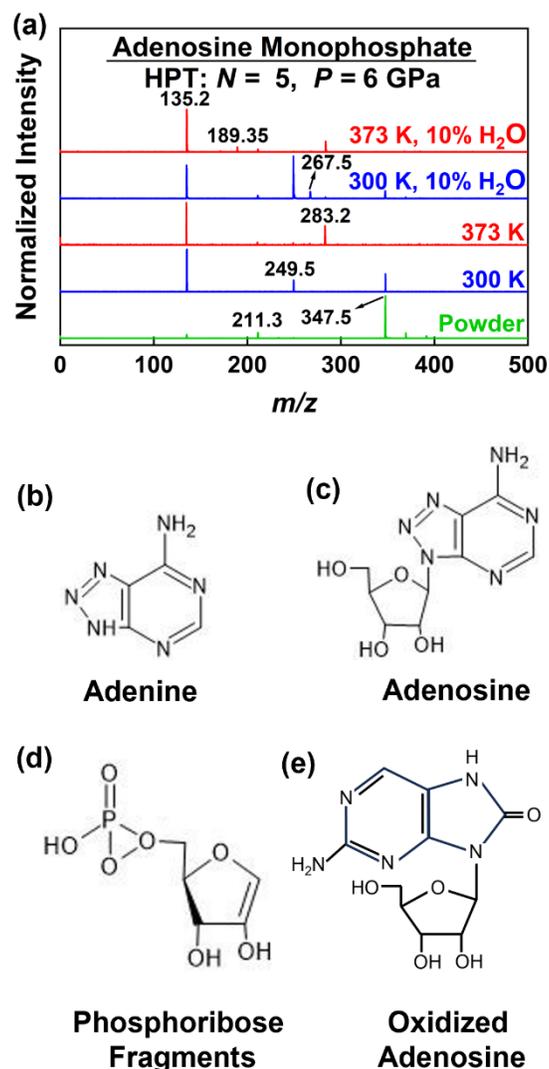

Fig. 6. (a) Matrix-assisted laser desorption/ionization time-of-flight mass spectroscopy profiles of adenosine monophosphate in dry and hydrated conditions before and following treatment through high-pressure torsion at ambient and boiling water temperatures. Molecular structure of (b) adenine, (c) adenosine (d) phosphoribose fragment and (e) oxidized adenosine.

SEM image analysis (Fig. 7) showed changes in the morphology of the powder after HPT processing at ambient temperature and a significant consolidation after HPT processing at 373 K. Consolidation by HPT is a common phenomenon for many kinds of powders that result from high processing pressure and severe plastic deformation (Edalati et al., 2024a).

Shear stress-strain curves in Fig. 8 indicate a strain hardening for AMP, with a slightly lower shear stress observed in hydrated samples. The shear stress was enhanced with an increase in shear strain and tended to be saturated at large strains, a response similar to that which occurs with metallic materials (Pippan et al., 2010). The steady shear strength of AMP treated at 373 K appeared to be larger than that of aluminum and slightly lower than that of copper processed at ambient temperature (Edalati et al., 2024b).

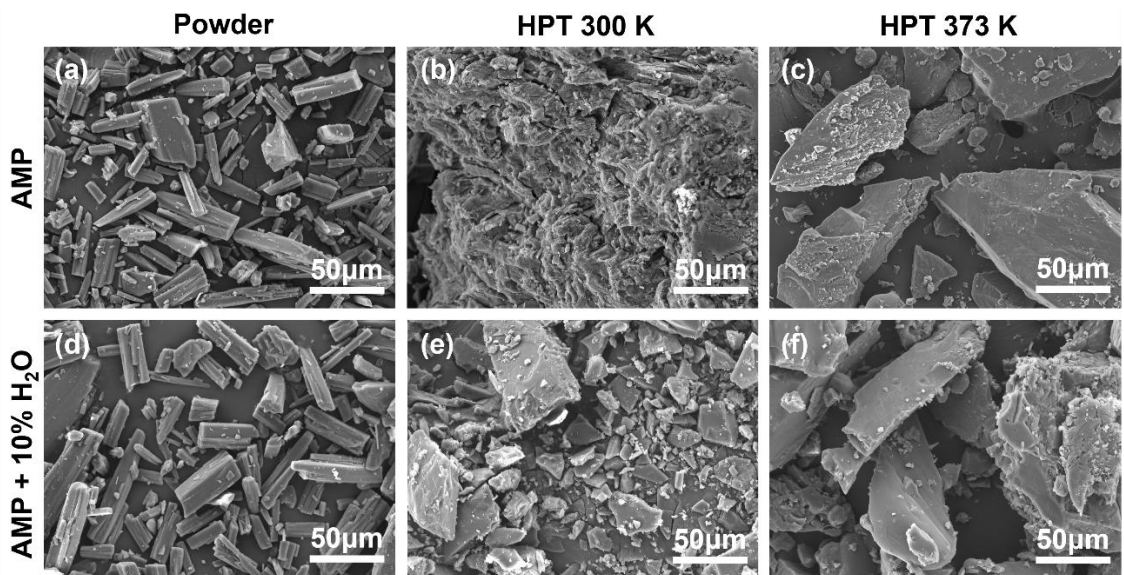

Fig. 7. Scanning electron microscopy images of adenosine monophosphate in (a-c) dry condition and (d-f) hydrated condition prior and following treatment through high-pressure torsion at ambient and boiling water temperatures.

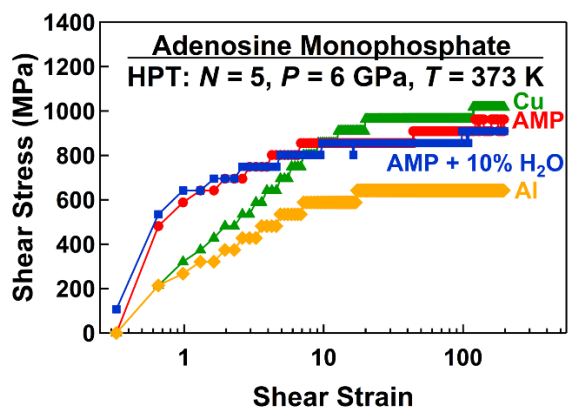

Fig. 8. Shear stress-strain curves, achieved by in situ torque measurements, for adenosine monophosphate in dry and hydrated conditions before and after treatment through high-pressure torsion at ambient and boiling water temperatures. Data for pure aluminum and copper processed at ambient temperature were included for comparison (Edalati et al., 2024b).

## 4. Discussion

The findings of the current investigation provide insight into the chemical transformations of AMP under HPT and simulate the extreme conditions associated with astronomical impacts. While no evidence of polymerization was detected, the formation of multiple AMP-derived fragments, including nucleobase-derived fragment adenine $C_5H_5N_5$, phosphoribose fragments $C_5H_9O_5P$, dehydrated adenosine $C_{10}H_{11}N_5O_3$, protonated adenosine $C_{10}H_{14}N_5O_4^+$ and oxidized adenosine $C_{10}H_{13}N_5O_5$ suggests that HPT promotes molecular breakdown and rearrangement under the tested conditions, expressed by the following equations.

$C_{10}H_{14}N_5O_7P \rightarrow C_5H_5N_5 + C_5H_9O_5P$ (3)

$C_{10}H_{14}N_5O_7P \rightarrow C_{10}H_{11}N_5O_3 + H_2O + H_3PO_4$ (4)

$C_{10}H_{14}N_5O_7P + 3H^+\ 1/2O_2\ (Air) \rightarrow C_{10}H_{14}N_5O_4^+ + H_3PO_4$ (5)

$$C_{10}H_{14}N_5O_7P + H_2O + 1/2 O_2 \text{ (Air)} \rightarrow C_{10}H_{13}N_5O_5 + H_3PO_4 \tag{6}$$

These transformations and the parameters that affected them are further addressed below.

The observed structural modifications can be linked to the severe pressure and strain imposed during HPT (Zhilyaev and Langdon, 2008; Edalati and Horita, 2016), which led to the breakage of specific molecular bonds. Raman spectroscopy and XRD revealed spectral peak-to-background intensity variations and peak shifts, which indicate structural disorder and defect formation (Kitajima, 1997; Oberdorfer, 2010; Čížek et al., 2019). Such changes suggest that HPT positively facilitates local molecular rearrangements but does not provide the necessary conditions for the formation of phosphodiester bonds essential for nucleotide polymerization. This finding is consistent with previous studies of high-pressure chemistry, which have demonstrated that additional catalytic or environmental factors, such as metal ions, minerals, or UV irradiation, are required to drive polymerization (Ferris et al., 1996; Orgel, 2004; Powner, 2009; Hazen and Sverjensky, 2010).

This study also addressed the impact of hydration on AMP stability. While hydrated samples exhibited similar fragmentation patterns to those of dry samples, slight differences in sample colors and characterization spectra suggest that water molecules influenced molecular rearrangement. Hydration has been shown to stabilize nucleotides under high-pressure conditions by facilitating hydrogen bonding and solvating reactive intermediates (Mast et al., 2013; Kitadai and Maruyama, 2018; Trigo-Rodriguez et al., 2019). However, the absence of significant polymerization in hydrated conditions suggests that water alone was insufficient to drive AMP assembly into larger nucleic acid structures. Future studies that incorporate mineral catalysts are likely required to elucidate the effects of hydration in prebiotic nucleotide transformations.

The temperature-dependent effects observed in this study are also noteworthy. Samples processed at 373 K exhibited more pronounced amorphization and fragmentation, as confirmed by XRD and MALDI-TOF mass spectrometry. The increase in molecular disorder at higher temperatures suggests that thermal energy enhanced molecular mobility, which increases the likelihood of chemical bond cleavage. However, despite these changes, no AMP oligomers or extended nucleotide chains were detected, which emphasizes the need for additional factors, such as catalytic agents, to facilitate polymerization (Ferris et al., 1996; Orgel, 2004; Powner, 2009; Hazen and Sverjensky, 2010).

Comparing the behavior of AMP to simpler biomolecules such as glycine provides additional context for interpreting these results. Prior HPT studies on amino acids have demonstrated molecular transformations without polymerization, with some amino acids forming secondary products such as alcohols (Edalati et al., 2022b). The observation that AMP undergoes fragmentation rather than polymerization under similar conditions suggests that more complex biomolecules may require additional environmental factors to facilitate bond formation. This supports the hypothesis that, while impact-generated pressures and strains contribute to molecular transformations, they must be complemented by additional prebiotic factors such as mineral templating or radiation to drive the formation of biologically relevant polymers (Ferris et al., 1996; Orgel, 2004; Powner, 2009; Hazen and Sverjensky, 2010).

Overall, these findings provide valuable insights into the potential role of astronomical impacts in prebiotic chemistry. While no polymerization of AMP into RNA precursors was observed by HPT processing, the decomposition and rearrangement of AMP into various organic fragments suggest that impact-generated conditions could have contributed to the chemical evolution of nucleotides on early Earth. The formation of nucleobase-derived fragments and phosphoribose units implies that mechanical forces and extreme pressures could facilitate the breakdown of biomolecules into simpler components, which may subsequently participate in secondary reactions that lead to the synthesis of more complex biopolymers under favorable conditions. However, the lack of phosphodiester bond formation under the tested conditions highlights the necessity of additional catalytic or environmental factors to drive the polymerization of nucleotides into RNA. These findings suggest that, while impact events can play a role in molecular transformations, the emergence of self-replicating genetic material likely required a more complex set of prebiotic conditions beyond mechanical stress alone. It should be noted that processing pure AMP is not a real simulation of early Earth conditions, and future research should explore the role of minerals, catalysts, and alternative pressure and strain regimes. Moreover, the combination of nucleotides with different phosphorylation modes is required to further understand the pathways that lead to RNA formation under prebiotic conditions. Moreover, extending such studies to other molecules, such as nucleoside cyclic phosphates, could further clarify the significance of astronomical impacts on prebiotic chemistry. This study hypothesizes that the HPT method could play an important role in elucidating the prebiotic chemistry of these biomolecules.

**Conclusions**

The present study demonstrates that high-pressure torsion (HPT) induces significant structural modifications in adenosine monophosphate (AMP) but does not lead to its polymerization under the tested conditions. The identification of various AMP-derived fragments, including nucleobase-derived adenine component $C_5H_5N_5$, phosphoribose fragment $C_5H_9O_5P$, dehydrated adenosine $C_{10}H_{11}N_5O_3$, protonated adenosine $C_{10}H_{14}N_5O_4^+$, and oxidized adenosine $C_{10}H_{13}N_5O_5$ highlights the role of mechanical stress in molecular rearrangements. These results support the hypothesis that astronomical impacts could have facilitated prebiotic transformations of nucleotides on early Earth. However, the absence of polymerization suggests that additional factors, such as catalytic minerals or UV radiation, are necessary for the formation of nucleic acids.

**Credit Authorship Contribution Statement**

All authors: Conceptualization, Methodology, Investigation, Validation, Writing – review & editing.

**Declaration of competing interest**

The authors declare no competing interests that could have influenced the results reported in the current article.

**Data availability**

All the data presented in this manuscript are available upon request from the corresponding author.

**Acknowledgments**

The author JHJ thanks the Q-Energy Innovator Fellowship of Kyushu University for a scholarship. This research received support from the Japan Society for the Promotion of Science via a Grant-in-Aid for Challenging Research (grant number: JP22K18737).